\documentstyle[12pt]{article}

\begin{document}
\begin{titlepage}

\begin{flushright}
SU-ITP-93-3\\ hepth@xxx/9302109\\ \today
\end{flushright}

\begin{center}

\baselineskip25pt
{\LARGE {\bf CHARGE QUANTIZATION OF\\
\vspace{.5cm}
AXION-DILATON BLACK HOLES}}

\vspace{1cm}
{\large {\bf Renata Kallosh}
\footnote{e-mail address: kallosh@physics.stanford.edu}
{\bf and Tom\'as Ort\'{\i}n}
\footnote{e-mail address: tomaso@slacvm}\\
\vspace{.5cm}
{\it Department of Physics}\\
{\it Stanford University}\\
{\it Stanford CA 94305, U.S.A.}}

\end{center}


\begin{abstract}
We present axion-dilaton black-hole and
multi-black-hole solutions of the low-energy string
effective action. Under $SL(2,R)$ electric-magnetic
duality rotations only the ``hair" (charges
and asymptotic values of the fields) of our solutions is
transformed.
The functional form of the solutions is duality-invariant.

 Axion-dilaton black holes with zero entropy and zero area
of the horizon form a family of stable particle-like objects,
which we call {\it holons}.

We study the quantization of the charges of these objects
and its compatibility with duality symmetry.
In general the spectrum
of black-hole solutions with quantized charges
is not invariant under $SL(2,R)$ but only
under $SL(2,Z)$ or one of its subgroups $\Gamma_{l}$.
Because of their transformation properties,
the asymptotic value of the
axion-dilaton field of a black hole may be associated
with the modular parameter $\tau$ of some complex
torus and the integer numbers $(n,m)$ that label its
quantized electric and magnetic charges may be
associated with winding numbers.
\end{abstract}

\end{titlepage}

\newpage

\baselineskip15pt
\pagestyle{plain}


{\bf 1.}  In this paper we are going to present
general four-dimensional static  axion-dilaton black hole solutions.
The  independent parameters characterizing such solutions
 are the black hole mass, electric and
 magnetic charges and  the values
of dilaton and axion fields at infinity. The dilaton and axion
charges
of this solutions are functions of these independent parameters.
Electric and magnetic charges of these black holes
will be quantized.

Our
solutions include zero area, zero entropy  axion-dilaton black holes,
carrying quantized electric and magnetic charges. In this sense such
objects
resemble elementary particles: we will call them {\it holons}.

The $SL(2,R)$ duality symmetry of the
classical equations of motion of the low-energy string
effective action \cite{kn:dual}, \cite{kn:modern} has
been used in \cite{kn:STW}, \cite{kn:Sendual} and
\cite{kn:TOM} to find new solutions with non-trivial
axion field. However, their final form  was not quite
satisfactory either because the axion and dilaton fields at infinity
were not
 independent parameters or
because the new solutions were expressed in
terms of the old charges and the parameters
of the transformation that have no physical meaning.
Here we present  the most general static solutions
that can be obtained in this way,
but expressed in terms of the new
charges only. Under a new duality
transformation only the ``hair" (charges and asymptotic
values of axion and dilaton fields, i.e. the boundary conditions)
will change in a well-established
way, while the functional form of the fields will
remain invariant.

We also obtained axion-dilaton multi-black hole solutions.
All fields in the
solutions are naturally built out of two complex harmonic
functions, which are real when the axion field identically
vanishes \cite{kn:Gibb}, \cite{kn:KLOPP}.

We will treat the $SL(2,R)$ transformations in a way
that will stress the analogies and
differences with the well-known $U(1)$ duality group of the
Einstein-Maxwell theory. In the Einstein-Maxwell case,
with the electric and magnetic charges $Q$, $P$,
one can build a two-component duality vector $(Q,P)$,
as explained in \cite{kn:Col}.
With axion and
dilaton fields, the canonical electric and magnetic
charges do not form $SL(2,R)$-duality vectors.
We will
present
combinations $(\tilde{q},p)$ of the canonical
charges  $Q, P$ and axion-dilaton asymptotic value $\lambda_{0}$ ,
 which do transform as duality vectors.

The issue of quantization of electric and magnetic charges
and the question  which  subgroup of $SL(2,R)$
will be compatible with the spectrum of black holes charges, can be
addressed
in the context of these families of black-hole solutions.
In the context of general string theory solutions
this problem has been studied by
Sen \cite{kn:Senquan} by embedding $U(1)$ into $SU(2)$ gauge group.
We will find the spectra of axion-dilaton black hole
charges with and without non-Abelian embedding, and will
classify the  symmetry groups of the spectra allowed by Dirac
quantization condition.

For $SL(2,Z)$-invariant spectra,
the $SL(2,R)$ duality vectors
$(\tilde{q},p)$, when properly normalized, take integer
values $(n,m)$  and label naturally the allowed states.
We will see that it is useful  to describe
the action of $SL(2,Z)$ on these spectra in terms of
$-1/\lambda_0= \tau$, the modular parameter of a complex
torus and the winding numbers $(n,m)$ of a
non-trivial homology cycle on this torus.

\vspace{0.8 cm}

{\bf 2.} Our conventions and our action are those of refs.
\cite{kn:KLOPP} and \cite{kn:TOM}, but here we
will use the complex scalar $\lambda=iz=a+ie^{-2\phi}$,
where $a$ is the axion field and $\phi$
is the dilaton field and
we will extend our analysis to a set of $U(1)$ vector
fields $A_{\mu}^{(n)}$, $n=1,2,\ldots,N$.

We find it convenient to define the
$SL(2,R)$-duals\footnote{The spacetime duals are
${}^{\star}F^{(n)\mu\nu}=\frac{1}{2\sqrt{-g}}
\epsilon^{\mu\nu\rho\sigma}F_{\rho\sigma}$, with
$\epsilon^{0123}=\epsilon_{0123}=+i $  .} to the
fields $F_{\mu\nu}^{(n)}= \partial_{\mu} A_{\nu}^{(n)}-
\partial_{\nu}
A_{\mu}^{(n)}$\ ,
\begin{equation}
\tilde{F}^{(n)}=e^{-2\phi}\,{}^{\star}F^{(n)}-iaF^{(n)}\; ,
\end{equation}
in terms of which the action reads
\begin{equation}\label{eq:action1}
S=
\frac{1}{16\pi}
\int d^{4}x\sqrt{-g}\biggl
\{-R+\frac{1}{2}\frac{\partial_{\mu}\lambda
\partial^{\mu}\overline{\lambda}}{({\mbox{Im}} \; \lambda)^{2}}
-\sum_{n=1}^{N}F^{(n)}_{\mu\nu}{}^{\star}
\tilde{F}^{(n)\mu\nu} \biggr \}\; .
\end{equation}
In terms of the component fields, we have
\begin{eqnarray}\label{eq:action2}
S & = &
\frac{1}{16\pi}
\int d^{4}x\sqrt{-g}\biggl \{-R+2(\partial\phi)^{2}
+\frac{1}{2}e^{4\phi}(\partial a)^{2}-
\nonumber \\ &&
-e^{-2\phi}\sum_{n=1}^{N}(F^{(n)})^{2}
+ia\sum_{n=1}^{N}F^{(n)}{}^{\star}F^{(n)}\biggr \}\; .
\end{eqnarray}

The advantage of using $\tilde{F}^{(n)}$ is that the
equations of motion imply the local existence of $N$
real vector potentials $\tilde{A}^{(n)}$ such that
\begin{equation}
\tilde{F}^{(n)}=i\, d\tilde{A}^{(n)}\; .
\end{equation}
The analogous equation
$F^{(n)}=dA^{(n)}$ is not a consequence of  equations
of motion but a consequence of the Bianchi identity.
If the time-like components $A_{t}^{(n)}$
play the role of electrostatic
potentials, then the $\tilde{A}^{(n)}_{t}$  will play the role of
magnetostatic potentials. As we will see
in the next section, the $SL(2,R)$ duality transformations
consist in the mixing of $A^{(n)}$ with $\tilde{A}^{(n)}$ and
of equations of motion with Bianchi identities
,as in the Einstein-Maxwell case.

Let us
stress at this point that the fields $\tilde{A}^{(n)}$ are
{\it not} dynamical fields of this theory. They
exist only on-shell. But they are particularly useful to
describe the field of magnetic monopoles from a strictly
classical point of view. The $A^{(n)}$'s
are the fields we are ultimately interested in, and
when we study the quantization problem, the description
provided by the $\tilde{A}^{(n)}$'s will not be sufficient.

Here we present two different kinds of static
solutions to the equations of motion of the action
(\ref{eq:action1}), (\ref{eq:action2}): spherical
black-hole solutions and multi-black-hole
solutions, both with nontrivial axion, dilaton and
$U(1)$ fields. All the previously
known solutions of these kinds (Schwarzschild, (multi-)
Reissner-Nordstr\"{o}m, the purely electric and magnetic
dilaton black holes of refs. \cite{kn:Gibb}, \cite{kn:GM}
and \cite{kn:GHS}, the electric-magnetic black holes
of refs. \cite{kn:KLOPP} and \cite{kn:TOM}, and the
axion-dilaton black holes of refs. \cite{kn:STW}
and \cite{kn:TOM}) are particular cases of them.

The static spherically symmetric black-hole solutions are
\begin{eqnarray}
ds^{2}                 & = &
e^{2U}dt^{2}-e^{-2U}dr^{2}-R^{2}d\Omega^{2}\; ,
\nonumber \\
\nonumber \\
\lambda(r)             & = &
\frac{\lambda_{0}r+\overline{\lambda}_{0}\Upsilon}{r
+\Upsilon}\; ,
\nonumber \\
\nonumber \\
A_{t}^{(n)}(r)         & = &
e^{\phi_{0}}R^{-2}[\Gamma^{(n)}(r+\Upsilon)+c.c]\; ,
\nonumber \\
\nonumber \\
\tilde{A}_{t}^{(n)}(r) & = &
-e^{\phi_{0}}R^{-2}[\Gamma^{(n)}(\lambda_{0}r+
\overline{\lambda}_{0} \Upsilon)+c.c]\; ,
\end{eqnarray}
where
\begin{eqnarray}
e^{2U}(r)  & = & R^{-2}(r-r_{+})(r-r_{-})\;,  \qquad
r_{\pm}    = M\pm r_{0}\; ,
\nonumber \\
R^{2}(r)   & = &  r^{2}-|\Upsilon|^{2}\; ,
\hspace{2,8 cm}r_{0}^{2}=M^{2}+|\Upsilon|^{2}
-4\sum_{n=1}^{N} |\Gamma^{(n)}|^{2}\; .
\end{eqnarray}
The  complex electromagnetic charge $\Gamma$,
axion-dilaton charge $\Upsilon$ and asymptotic values of the fields
 are defined in the Appendix.
The singularity is hidden under a horizon if $r_{0}^{2}>
0$, and it is hidden or coincides with it (but still is
invisible for external observers) if $r_{0}=0$.
The  conditions
$r_{0}^{2} \geq 0$
 and $M\geq |\Upsilon|$ can be related to supersymmetry bounds
\cite{kn:KLOPP}, \cite{kn:TOM}, \cite{kn:FUTURE}. All solutions given
above
have the entropy
\begin{equation}
S= \pi (r_{+}^2 - |\Upsilon|^{2}) \ .
\end{equation}
When all  supersymmetric bounds are saturated, i.e.
$r_{+}=M=|\Upsilon|$,
the objects described by this solution have
 zero area of the horizon and vanishing entropy.  They may be
considered as
the ground states of the theory and we will call them
 {\it
holons}. A detailed discussion of
thermodynamics of all solutions and their
physical interpretation will be given in
\cite{kn:FUTURE}.

Our second
kind of solutions describe  axion-dilaton extreme multi-black-hole
solutions.
 The fields are
\begin{eqnarray}
ds^{2}           & = & e^{2U}dt^{2}-e^{-2U}d\vec{x}^{2}\;
, \hspace{1cm}
                                e^{-2U}(\vec{x})=2 \;{\mbox{Im}}\;
                             ({\cal H}_{1}(\vec{x})\;
                                 \overline{{\cal
                                H}}_{2}(\vec{x}))\; ,
                                    \nonumber \\
\nonumber \\
\lambda(\vec{x}) & = & \frac{{\cal H}_{1}(\vec{x})}{{\cal
                         H}_{2}(\vec{x})}\; ,
                        \nonumber \\
\nonumber \\
A_{t}^{(n)}(\vec{x})       & = & e^{2U}(k^{(n)}{\cal
                                H}_{2}(\vec{x})+c.c)\; ,
                        \nonumber \\
\nonumber \\
\tilde{A}^{(n)}_{t}(\vec{x}) & = & -e^{2U}(k^{(n)}{\cal
                                 H}_{1}(\vec{x})+c.c)\; ,
\end{eqnarray}
where ${\cal H}_{1}(\vec{x}), {\cal H}_{2}(\vec{x})$
are two complex harmonic functions
\begin{eqnarray}
{\cal H}_{1}(\vec{x}) & = & \frac{e^{\phi_{0}}}{\sqrt{2}}
                            \{\lambda_{0}+\sum_{i=1}^{I}
                            \frac{\lambda_{0}M_{i}+
                            \overline{\lambda}_{0}
                            \Upsilon_{i}}{|\vec{x}
                            -\vec{x}_{i}|}\}\; ,
                                      \nonumber \\
\nonumber \\
{\cal H}_{2}(\vec{x}) & = & \frac{e^{\phi_{0}}}{\sqrt{2}}
                            \{1+\sum_{i=1}^{I}\frac{M_{i}+
                            \Upsilon_{i}}{|\vec{x}
                            -\vec{x}_{i}|}\}\; .
\end{eqnarray}

The horizon of the $i$-th (extreme) black hole is at
$\vec{x}_{i}$ (in these
isotropic coordinates the horizons look like single
points), and has mass $M_{i}$, electromagnetic
charge $\Gamma_{i}$, etc. , as can be seen by using the
definitions in the Appendix
in the limit $|\vec{x}-\vec{x}_{i}|\rightarrow\infty$.
Charges without
label are total charges. The constants $k^{(n)}$ are
\begin{equation}
k^{(n)}=
-\frac{1}{\sqrt{2}}
\biggl (\frac{\Gamma^{(n)}M+\overline{\Gamma^{(n)}
\Upsilon}}{M^{2}-|\Upsilon|^{2}}\biggr )\; .
\end{equation}
The consistency of the solution requires for every $i$
\begin{equation}
k^{(n)}_{i}=k^{(n)}\; ,
\hspace{1cm}
Arg(\Upsilon_{i})=Arg(\Upsilon)\; .
\end{equation}
Finally, for each $i$ and also for the total charges, the
supersymmetric Bogomolny bound is saturated:
\begin{equation}
M^{2}+|\Upsilon|^{2}-
4\sum_{n=1}^{N}|\Gamma^{(n)}|^{2}=0\; .
\end{equation}

For a single $U(1)$ vector field
 the extreme solution simplifies to ($k\equiv k^{1}$, $\Gamma \equiv
\Gamma^{1}$)
\begin{equation}
M^{2}=|\Upsilon|^{2}\ , \quad k
 =  -\frac{\Gamma}{2 \sqrt{2} M}
\; .\end{equation}
As a consequence
of all the identities obeyed by the charges, it is
possible to derive the following expression of equilibrium
of forces between two extreme black holes \cite{kn:FUTURE}:
\begin{equation}\label{eq:equiforce}
M_{1}M_{2}+\Sigma_{1}\Sigma_{2}+\Delta_{1}\Delta_{2}=
Q_{1}Q_{2}+
P_{1}P_{2}\; .
\end{equation}

\vspace{0.8cm}

{\bf 3.} $SL(2,R)$ duality acts on the fields of our
theory as follows:
\begin{eqnarray}
A^{(n)\prime}(x)         & = & \hskip 0.4 cm \delta
A(x)-\gamma\tilde{A}(x)\; ,
                                            \nonumber \\
\tilde{A}^{(n)\prime}(x) & = & -\beta A(x)+
\alpha\tilde{A}(x)\; , \label{eq:trans}
\end{eqnarray}
where $\alpha, \beta, \gamma$ and $\delta$
are the elements of a $SL(2,R)$ matrix
\begin{equation}
R=
\left(
\begin{array}{cc}
\alpha & \beta  \\
\gamma & \delta \\
\end{array}
\right)\; .
\end{equation}
Notice that, since the $\tilde{A}^{(n)}$'s
are not independent fields, the consistency of eqs.
(\ref{eq:trans})
implies the transformation law of $\lambda$:
\begin{equation}
\lambda^{\prime}(x)=\frac{\alpha\lambda(x)+
\beta}{\gamma\lambda(x)+\delta}\; .
\end{equation}
With no dilaton nor axion ($\lambda=i$), our theory coincides with
Einstein-Maxwell theory. In this case $\tilde{F}={}^{\star}F$ and the
consistency of eqs. (\ref{eq:trans}) would imply that
$R$ is a $SO(2)$ matrix, the
duality group being just $U(1)$.

As in the Einstein-Maxwell
case, the duality transformations  (\ref{eq:trans})
rotate continuously equations of motion into Bianchi
identities. In both cases the equations of motion are
invariant under duality but the actions are not.

Many objects in this theory have well-defined
transformation properties under duality, i.e. they
transform according to some representation of $SL(2,R)$,
usually the vector representation or its dual. We can
speak about duality vectors (pairs that transform
with $R$) and duality forms (pairs that transform with
$R^{-1}$), in the language of ref. \cite{kn:Col}.

Using the definitions of the charges in the Appendix,
and eqs. (\ref{eq:trans}), which can be written as
\begin{equation}
(A^{(n)\prime}(x), \tilde{A}^{(n)\prime}(x))
=
(A^{(n)}(x),
\tilde{A}^{(n)}(x))      (R^{-1})^{T}\; ,
\end{equation}
we get the following
transformation laws for the ``hair" of any field
configuration and, in particular, of our solutions:
\begin{equation}
\left(
\begin{array}{c}
\tilde{q}^{\prime} \\
p^{\prime}         \\
\end{array}
\right)
=R
\left(
\begin{array}{c}
\tilde{q} \\
p        \\
\end{array}
\right)\; ,
\hspace{2cm}
(q^{\prime}, \tilde{p}^{\prime})=
(q, \tilde{p})(R^{-1})^{T}\; ,
\end{equation}
\begin{equation}
\lambda_{0}^{\prime}  =
\frac{\alpha\lambda_{0}+\beta}{\gamma
\lambda_{0}+\delta}\; ,
\hspace{2cm}
\Upsilon^{\prime}   =
e^{-2iArg(\gamma\lambda_{0}+\delta)}\Upsilon\; ,
\end{equation}
\begin{equation}
\Gamma^{\prime}=e^{+iArg(\gamma\lambda_{0}+\delta)}
\Gamma\; . \label{eq:transhair}
\end{equation}
The pairs $(q^{(n)},\tilde{p}^{(n)})$ and
$(A^{(n)}(x),\tilde{A}^{(n)}(x))$
are duality forms and the pairs
$(\tilde{q}^{(n)},p^{(n)})$ are
duality vectors. It should be possible to
express any observable property of a system of
two dyons in terms of $SL(2,R)$-invariant bilinears of
their charges. The first invariant is the scalar product
of two duality vectors, which is equivalent to
the action of a duality form on a duality vector:
\begin{equation}
(q_{1},\tilde{p}_{1})
\left(
\begin{array}{c}
\tilde{q}_{2} \\
p_{2}         \\
\end{array}
\right)
= q_{1}\tilde{q}_{2}+\tilde{p}_{1}p_{2}
=Q_{1}Q_{2}+P_{1}P_{2}\; .
\end{equation}
This is symmetric and appears in the expression of the
``Coulomb" force between two static
dyons (see eq. (\ref{eq:equiforce})). The second
invariant is the exterior product of two duality vectors
or forms, which corresponds to the simplectic form
\begin{equation}
S=\left(\begin{array}{cc}0&1\\-1&0\\
\end{array} \right)\; .
\end{equation}
It is obviously invariant under $Sp(2,R)=SL(2,R)$:
\begin{equation}
(\tilde{q}_{1},p_{1})S
\left(\begin{array}{c}\tilde{q}_{2} \\
p_{2}\end{array} \right) =
(q_{1},\tilde{p}_{1})S
\left(\begin{array}{c}q_{2}\\
\tilde{p}_{2}\end{array} \right) =
Q_{1}P_{2}-Q_{2}P_{1}\; .
\end{equation}
This is antisymmetric and will appear
in the quantization condition.

Now, using eqs. (\ref{eq:trans}) in the
solutions presented in the previous section, one can
check that the effect of a $SL(2,R)$ transformation on
them is equivalent to a $SL(2,R)$ transformation of
the ``hair" according to eq. (\ref{eq:transhair}).
 The transformation of the fields at
any point of the spacetime is equivalent to the
transformation of the boundary conditions only.

\vspace{0.8cm}

{\bf 4.}
It is well known that, in general, the introduction of
magnetic charges is not compatible with quantum mechanics.
In the case of two dyons interacting through their
electric and magnetic monopole fields, the quantization
of the system will be consistent only if the charges obey
Dirac-Schwinger-Zwanziger's condition \cite{kn:DSZ}
\begin{equation}\label{eq:DSZ}
Q_{1}P_{2}-Q_{2}P_{1}=\frac{n}{2}\; ,
\end{equation}
where $n$ is any integer number. Naively eq.
(\ref{eq:DSZ}) can be
obtained by quantization of the angular momentum
of the electromagnetic field. This condition ensures the
single-valuedness of the system's wave function, i. e.
that the string singularity in the vector potential can
not be detected in an Aharonov-Bohm-like experiment. On
the other hand, eq. (\ref{eq:DSZ}) is invariant under
the $U(1)$ duality rotations of Maxwell's theory whose
preservation was the main idea behind Dirac's work.

We have described our solutions in terms of the
``dual potentials" $\tilde{A}_{\mu}^{(n)}$,
but the fundamental objects in quantum mechanics
are the potentials $A_{\mu}^{(n)}$.
Their space-like components $A_{i}^{(n)}$ can be
expressed in terms of the magnetostatic potentials
$\tilde{A}^{(n)}_{0}$.
In presence of magnetic charges they will
exhibit string singularities or will
have to be defined in different gauges in
different regions of the space \cite{kn:WY}.
To achieve consistency, conditions
like eq. (\ref{eq:DSZ}) will have to be satisfied.

If we want to study  a quantum system
consisting of two of our dyons moving in
each other's very distant field\footnote{For
the sake of simplicity we consider a single $U(1)$
field in this section.}, only the asymptotic behavior
of the potentials matters and the black hole
nature of our dyons becomes
irrelevant. But this is just the same of
dyons in Minkowski space, and the same string singularities
must be present. Then, the consistency condition
(calculated with our charge conventions given in the
Appendix) takes the standard form of  eq. (\ref{eq:DSZ}).

In terms of our duality vectors and
duality forms the quantization condition is
\begin{equation}\label{eq:quant}
\tilde{q}_{1}p_{2}-\tilde{q}_{2}p_{1}=
q_{1}\tilde{p}_{2}-q_{2}\tilde{p}_{1}=\frac{n}{2}\; ,
\end{equation}
that is, the exterior product \cite{kn:Col} of the charge
duality vectors or forms of the two dyons is a half-integer.
This statement is invariant under the full $SL(2,R)$.

However, these
conditions only restrict the product of the charges and
do not tell us anything about what the individual charges
might be. To go further we need additional information
(basically the spectrum of electric charges allowed by gauge
invariance).
In ref. \cite{kn:Sendual} the information
about the spectrum of electrically charged states was
taken from string theory.
Here we will study two cases: with and without non-Abelian
embedding.
In the first case we essentially follow the approach of ref.
\cite{kn:Sendual},
embedding our $U(1)$ vector field in $SU(2)$:
$A_{\mu}= B_{3\mu}$.
Asymptotically our theory is equivalent to a spontaneously
broken $SU(2)$ theory with a CP-violating term of the form
$\sim \theta F{}^{\star}F$. The relation between $g$, the
coupling constant\footnote{Only in the framework of a
non-Abelian theory is this identification possible, since
in the Abelian theory the term $F^{2}$ does not contain
couplings between charged particles.} and $\theta$ of
this theory and the asymptotic values of the dilaton
field $\phi_{0}$ and axion field $a_{0}$ are given by
\begin{equation}
e^{-2\phi_{0}}=\frac{4\pi}{g^{2}}\; ,
\hspace{2cm}
a_{0}=\frac{\theta}{2\pi}\; .
\end{equation}

In this theory there is a natural unit of charge $c$
which with  our charge conventions is
\begin{equation}
c=\frac{g}{\sqrt{4\pi}}=e^{+\phi_{0}}\; .
\end{equation}
When
$a_{0}=0$ the electric charge of a dyon is an integer
multiple of this elementary charge $Q=nc$.
However, Witten showed in ref. \cite{kn:W} that in presence
of the CP-violating
$\theta$  and magnetic charge $P$, the value of $Q$ is corrected by a
term proportional to $\theta$. In  our case the corrected charge is
given by
\begin{equation}\label{eq:Witten}
Q=nc-\frac{\theta}{2\pi} c^{2}P=
e^{+\phi_{0}}(n-a_{0}e^{+\phi_{0}}P)\; ,
\end{equation}
which means that $\tilde{q}=n$. Using
eq. (\ref{eq:quant}) for a purely electric state and state
with magnetic charge we see that the allowed values for $P$
are $P=m/2c$, that is, $p=m/2$. However, since this
theory admits particles with elementary charge $c/2$, the
half-integer values of $p$ have to be excluded
in case such particles exist (all the
magnetic charges have  be multiples of the elementary
charge of 't Hooft-Polyakov's monopole). Hence,
if we denote each dyon state by the duality vector
$(\tilde{q},p)$, then the spectrum will be given by
\begin{equation}\label{eq:specsu2}
Q+iP=e^{+\phi_{0}}(n-\overline{\lambda_{0}}m)\; ,
\hspace{2cm}
(\tilde{q},p)  = (n,m)\; , \hspace{1cm}n,m\in Z\; .
\end{equation}

This is a one-dimensional version of the spectrum of ref.
\cite{kn:Sendual}.
It is clear that
the form of the spectrum in eq. (\ref{eq:specsu2}) is not
respected by arbitrary $SL(2,R)$ transformations, but only
when $\alpha,\beta,\gamma,\delta$ are all integers. Then, the
spectrum (\ref{eq:specsu2}) is invariant under $SL(2,Z)$
(the transformations act on the states as permutations).
In particular, when $a_{0}=0$, the spectrum is
invariant under the
$Z_{2}$ subgroup $Q\rightarrow P, \; P\rightarrow -Q$.
This is possible because $c=e^{\phi_{0}}$
does change according to $c\rightarrow c^{-1}$
under the former transformation. This invariance
is not present in the Maxwell-Dirac
case $Q=ne,\, P=me^{-1}$
because $e$ does not change under the $U(1)$ duality
group.

How do our results depend on the embedding of
$U(1)$ in a non-Abelian\footnote{This
case is particularly important for supersymmetric
black holes \cite{kn:KLOPP},
\cite{kn:FUTURE}. It was also stressed by Schwarz
\cite{kn:modern} that it is not known how to maintain the $SL(2,R)$
symmetry
when non-Abelian vector fields and/or charged fields are
introduced.} group? The most important
difference would be the absence of a natural unit of
charge. Once a unit of charge $c$ is defined, because of
$U(1)$ gauge symmetry, the electric charge is quantized
in integer multiples of it. In general the spectrum will
break completely the duality invariance as
in the Einstein-Maxwell case, since the duality group
does not act on $c$. A more interesting starting
point would be to assume that the elementary charge
is a factor $\xi$ times $e^{+\phi_{0}}$, $c=\xi e^{+\phi_{0}}$, and
so the
electric charge is again given by eq. (\ref{eq:Witten}).
Using the most general magnetic charges allowed
by eq. (\ref{eq:DSZ}) we are led to the spectrum
\begin{equation}\label{eq:specxi}
(\tilde{q},p)=(n\xi\, ,\, \frac{m}{2\xi})\; ,
\hspace{.5cm}n,m\in Z\; .
\end{equation}

This spectrum includes states $\tilde{q}p=\frac{1}{2}
(\mbox{mod}\, Z)$ that behave as fermions \cite{kn:Gold}, while
(\ref{eq:specsu2}) does not.
Again, for general $\xi$, the spectrum (\ref{eq:specxi})
breaks $SL(2,R)$ completely because, in principle, $SL(2,R)$
does not act on $\xi$. Naively, the introduction of $\xi$
seems to be completely equivalent to a rescaling
of $e^{+\phi_{0}}$ such that $e^{+\phi_{0}^{\prime}}=\xi
e^{+\phi_{0}}$. However, there is no
reason why $\xi
e^{+\phi_{0}}$
should transform like $e^{\phi_{0}^{\prime}}$. We can consider both
possibilities ($\xi$ invariant and $\xi$ transforming under $SL(2,R)$
in the
right way) by using explicitly a $SL(2,R)$ invariant $\xi$. The case
in which
it is legitimate to absorb $\xi$ in a rescaling of $e^{+\phi_{0}}$,
that is, the
case in which the elementary charge $c$ transforms as
$e^{+\phi_{0}}$,
will be just the case $\xi=1$ in what follows.

If $2\xi^{2}=l\in Z$, it is not hard to check that
(\ref{eq:specxi}) is still invariant under $\Gamma_{l}$,
the subgroup of matrices of $SL(2,Z)$ with $\beta=0\,
(\mbox{mod}\ l)$.  The case $l=1$ is special, since
$\Gamma_{1}=SL(2,Z)$. This means that for $\xi=1/\sqrt{2}$
the spectrum  (\ref{eq:specxi}) has the maximum amount of
symmetry.

When $l\neq 1$, the spectrum has less symmetry than
$\Gamma_{1}=SL(2,Z)$. Nevertheless, if we restrict
$m$ to be a multiple of $l$, $m=rl$, projecting out of the
spectrum the other states, we recover full $SL(2,Z)$
invariance and the spectrum will have the form
\begin{equation}\label{eq:specxi2}
(\tilde{q},p)=\sqrt{\frac{l}{2}}\; (n,r)\;
,\hspace{.5cm}n,r,l\in Z\; .
\end{equation}

The spectrum
will contain fermionic states if $l$ is odd. The first
spectrum (\ref{eq:specsu2}) which results from the embedding in
$SU(2)$ is
the case $l=2$ of (\ref{eq:specxi2}). On the
other hand, (\ref{eq:specxi2}) seems to be the most general
form of a spectrum invariant under the full $SL(2,Z)$.
The lesson is that no part of the original $SL(2,R)$ duality
symmetry will be respected by the spectrum after quantization if
the elementary electric and magnetic charges (which are
essentially the inverse of each other) do not have
the right transformation properties under duality.
However, if they do have special transformation properties under
duality,
the largest subgroup of the classical symmetry, which is compatible
with quantization, is $SL(2,Z)$.

The solutions presented in this paper provide explicit
realizations of the states in all these spectra.

\vspace{.8cm}

{\bf 5.}
A useful way of thinking  of the transformation rules of
the hair of our black holes after quantization is the
following. It is always possible to
interpret $\lambda_{0}$ as the modular parameter of
some complex torus, transforming under the
modular group as described in eq. (\ref{eq:transhair}).
But it is also possible to consider $-1/\lambda_{0}$
as the modular parameter $\tau$ of another (conformally
equivalent) torus. This means that we have
chosen a canonical basis of homology cycles $A,B$ on the torus and we
have
normalized the unique Abelian differential $\omega(z)dz$ by
\begin{equation}
\oint_{A}\omega(z)dz =1\; ,
\end{equation}
and when we integrate over the $B$ cycle we get
\begin{equation}
\oint_{B}\omega(z)dz=\tau=-1/\lambda_{0}\; .
\end{equation}
Under the $SL(2,Z)$ transformation  (\ref{eq:transhair}),
$\tau$ transforms as
\begin{equation}
\tau^{\prime}=\frac{\delta\tau-\beta}{\gamma\tau+\alpha}\; ,
\end{equation}
which corresponds to the following transformation of the
homology basis:
\begin{equation}
\left(
\begin{array}{c}
B ^{\prime} \\
A^{\prime} \\
\end{array}
\right)
=
\left(
\begin{array}{cc}
\delta        & -\beta     \\
 -\gamma & \alpha   \\
\end{array}
\right)
\left(
\begin{array}{c}
B  \\
A \\
\end{array}
\right)\; . \label{eq:basis}
\end{equation}

Now, if we have any homology cycle $C$ on the surface
of the torus, it can be expressed in terms of the
basis $A,B$ as $C=mA+nB$ where the
equal sign means "homologically equivalent". The integers
$n$ and $m$ count the number of times the cycle $C$
wraps around the cycles $B$ and $A$ respectively, and are
called {\it winding numbers}. In terms of the
transformed basis $A^{\prime},B^{\prime}$, we will have
$C=m^{\prime}A^{\prime}+n^{\prime}B^{\prime}$, and the new
winding numbers will be related with the old ones by
\begin{equation}
\left(
\begin{array}{c}
n^{\prime} \\
m^{\prime} \\
\end{array}
\right)
=
\left(
\begin{array}{cc}
\alpha & \beta     \\
\gamma & \delta    \\
\end{array}
\right)
\left(
\begin{array}{c}
n  \\
m  \\
\end{array}
\right)\; .
\end{equation}

For any $SL(2,Z)$-invariant spectrum of the
form (\ref{eq:specxi2}) this is exactly
the way our integer quantum numbers $(n,m)$ transform.
Then, in this framework, we can state our result in this
way: to any static spherically
symmetric dilaton-axion black hole with the hair
$M, \;Q(n,m),\; P(m),\; \lambda_{0}$, we can associate a
complex torus of modular parameter $\tau=-1/\lambda_{0}$,
and a homology cycle $C$ of winding numbers $(n,m)$ on it.

We hope to present a detailed discussion of
different aspects of quantized
axion-dilaton  black holes in a subsequent publication
\cite{kn:FUTURE}.

\vspace{.8cm}

{\large {\bf Acknowledgements}}

We would like to thank M. Dine, L. Dixon, J. Harvey, M.
Peskin, A. Sen, J. Schwarz, A. Strominger and L. Susskind
for most useful discussions. We are especially grateful
to A. Linde for the help in setting up the issue of
black hole charge quantization and illuminating
discussions. The work of R.K. has been
supported by the NSF grant PHY-8612280 and in part by
Stanford University. The work of T.O. has been supported
by a Spanish Government MEC postdoctoral grant.

\vspace{.8cm}

{\large{\bf Appendix: Definitions of the charges}}

We define the complex charges
in terms of the asymptotic behavior ($r\rightarrow
\infty$) of the different complex fields
\begin{eqnarray}
g_{tt} & \sim  & 1-\frac{2M}{r}\; ,
\hspace{3cm}
\lambda \sim \lambda_{0}-ie^{-2\phi_{0}}
\frac{2\Upsilon}{r}\; , \nonumber \\
F_{tr}  & \sim
&\frac{e^{+\phi_{0}}Q}{r^{2}}=
\frac{q}{r^{2}}\; . \hspace{2cm}
{}^{\star}F_{tr}  \sim \frac{i
e^{+\phi_{0}}P}{r^{2}}= \frac{ip}{r^{2}}\; .
\end{eqnarray}
The relation between these charges and the canonically
normalized $SU(2)$ charges of ref. \cite{kn:W} is
\begin{equation}
Q=\frac{Q^{can}}{\sqrt{4\pi}}\; ,\hspace{1cm}
P=\frac{P^{can}}{\sqrt{4\pi}}\; .
\end{equation}
The real axion ($\Delta$), dilaton ($\Sigma$), electric
($Q$) and magnetic ($P$) charges, and the
asymptotic values of the
axion ($a_{0}$) and dilaton ($\phi_{0}$) are
\begin{equation}
\Upsilon  = \Sigma-i\Delta,
\hspace{1cm}
\Gamma   = \frac{1}{2}(Q+iP),
\hspace{1cm}
\lambda_{0}  =  a_{0}+ie^{-2\phi_{0}}.
\end{equation}
The definition of the charges
associated to $\tilde{F}$ are
\begin{equation}
\tilde{F}_{tr}       \sim
\frac{i\tilde{p}}{r^{2}}\; , \hspace{1cm}
{}^{\star}\tilde{F}_{tr}\sim
\frac{\tilde{q}}{r^{2}}\; ,
\end{equation}
so we have for the components of the
duality vectors and forms
\begin{eqnarray}
q & = & e^{+\phi_{0}}Q\; ,
\hspace{3.3cm}
\tilde{q} = (e^{-\phi_{0}}Q+a_{0}e^{+\phi_{0}}P)\; ,
\\ \nonumber
\tilde{p} & = & (e^{-\phi_{0}}P-a_{0}e^{+\phi_{0}}Q)\; ,
\hspace{1cm}
p = e^{+\phi_{0}}P\;  .
\end{eqnarray}
In every single black hole in our solutions the charge of
the complex scalar is related to the $N$ gauge charges by
\begin{equation}
\Upsilon=- \frac{2
}{M}\, \sum_{n=1}^{n=N} (\overline{\Gamma}_{n})^{2}\; .
\end{equation}

\end{document}